\documentclass[sigconf]{acmart}
\AtBeginDocument{%
  }

\copyrightyear{2026}
\acmYear{2026}
\setcopyright{cc}
\setcctype{by-nc-nd}
\acmConference[DAC '26]{63rd ACM/IEEE Design Automation Conference}{July 26--29, 2026}{Long Beach, CA, USA}
\acmBooktitle{63rd ACM/IEEE Design Automation Conference (DAC '26), July 26--29, 2026, Long Beach, CA, USA}
\acmDOI{10.1145/3770743.3804164}
\acmISBN{979-8-4007-2254-7/2026/07}

\usepackage{multirow}
\usepackage{subfigure}
\begin{document}

\title{Spike-PTSD: A Bio-Plausible Adversarial Example Attack on Spiking Neural Networks via PTSD-Inspired Spike Scaling}

\author{Lingxin Jin}
\email{jinlx@std.uestc.edu.cn}
\orcid{0000-0002-4120-5841}
\affiliation{%
  \institution{University of Electronic Science and Technology}
  \city{Chengdu}
  \country{China}
}
\affiliation{%
  \institution{Khalifa University}
  \city{Abu Dhabi}
  \country{The United Arab Emirates}
}

\author{Wei Jiang}
\email{weijiang@uestc.edu.cn}
\orcid{0000-0001-6181-3900}
\affiliation{%
  \institution{University of Electronic Science and Technology}
  \city{Chengdu}
  \country{China}
}

\author{Maregu Assefa Habtie}
\email{maregu.habtie@ku.ac.ae}
\orcid{0000-0003-2815-7993}
\affiliation{%
  \institution{Khalifa University}
  \city{Abu Dhabi}
  \country{The United Arab Emirates}
}

\author{Letian Chen}
\email{chenletian03@std.uestc.edu.cn}
\orcid{0009-0008-9678-1636}
\affiliation{%
  \institution{University of Electronic Science and Technology}
  \city{Chengdu}
  \country{China}
}

\author{Jinyu Zhan}
\email{zhanjy@uestc.edu.cn}
\orcid{0000-0002-0214-7124}
\affiliation{%
  \institution{University of Electronic Science and Technology}
  \city{Chengdu}
  \country{China}
}

\author{Xingzhi Zhou}
\email{xingzhizhou@std.uestc.edu.cn}
\orcid{0009-0001-9496-0285}
\affiliation{%
  \institution{University of Electronic Science and Technology}
  \city{Chengdu}
  \country{China}
}

\author{Lin Zuo}
\email{linzuo@uestc.edu.cn}
\orcid{0000-0002-1598-1249}
\affiliation{%
  \institution{University of Electronic Science and Technology}
  \city{Chengdu}
  \country{China}
}

\author{Naoufel Werghi}
\email{naoufel.werghi@ku.ac.ae}
\orcid{0000-0002-5542-448X}
\affiliation{%
  \institution{Khalifa University}
  \city{Abu Dhabi}
  \country{The United Arab Emirates}
}

\renewcommand{\shortauthors}{Jin et al.}

\begin{abstract}
Spiking Neural Networks (SNNs) are energy-efficient and biologically plausible, ideal for embedded and security-critical systems, yet their adversarial robustness remains open. Existing adversarial attacks often overlook SNNs' bio-plausible dynamics. We propose \textbf{Spike-PTSD}, a biologically inspired adversarial attack framework modeled on abnormal neural firing in Post-Traumatic Stress Disorder (PTSD). It localizes decision-critical layers, selects neurons via hyper/hypoactivation signatures, and optimizes adversarial examples with dual objectives. Across six datasets, three encoding types, and four models, Spike-PTSD achieves over 99\% success rates, systematically compromising SNN robustness. Code: https://github.com/bluefier/Spike-PTSD.
\end{abstract}



\maketitle

\section{Introduction}
Spiking Neural Networks (SNNs), also known as the third-generation neural networks, simulate brain-like information processing through discrete spike events \cite{third_generation_spike,time_spike}. Unlike conventional Artificial Neural Networks (ANNs) based on continuous computation (\cite{VGG, ResNet-He, LSTM, BERT, Transformer, ViT}), SNNs achieve higher biological plausibility and lower energy consumption, showing strong potential in embedded systems, edge intelligence, and robotics \cite{intellegent_use}. Neuromorphic hardware such as Intel Loihi \cite{Loihi}, IBM TrueNorth \cite{TrueNorth}, and SpiNNaker \cite{Spinnaker} further enables on-chip spike computation with ultra-low power and low latency. However, these deployments introduce new system-level attack surfaces. In neuromorphic platforms, the event-based interface between sensors (e.g., DVS cameras \cite{DVS-Camera}) and neuromorphic cores can be exploited to inject abnormal spike streams \cite{hardware_trojan}. Meanwhile, hardware non-idealities such as quantized membrane voltages or spike-timing jitter amplify these effects \cite{trustworthy_neuro_comp, 28nm}. Hardware Trojans and bit flip attacks can further compromise neuromorphic integrity \cite{hardware_trojan, ETBT}. Thus, algorithmic vulnerabilities in SNNs may propagate to chip-level failures, threatening safety-critical applications and emphasizing the need for design-time validation and hardware trust assurance \cite{hardware_security}.

As neuromorphic hardware becomes increasingly integrated into real-world systems, such adversarial behaviors pose tangible risks to hardware-level trust and reliability, highlighting the internal vulnerabilities of SNN-based architectures \cite{hardware_security}. Existing works expose these vulnerabilities through various Adversarial Example Attacks (AEAs) \cite{SNN_robustness_anlaysis,SPA,RGA,HART,PDSG}. Early methods relied on equivalent ANNs to circumvent the non-differentiability of spike backpropagation \cite{SNN_robustness_anlaysis}; later methods such as RGA \cite{RGA} and HART \cite{HART} introduced surrogate gradients for direct optimization in SNNs; and PDSG \cite{PDSG} further adapted gradients to membrane potential dynamics. Despite their success, these methods overlook the bio-plausible mechanisms that govern spike generation and propagation. Consequently, they generalize poorly on dynamic (or neuromorphic) datasets and degrade sharply under targeted attack settings.
\begin{figure}[!t]
    \centering
    \includegraphics[width=0.99\columnwidth]{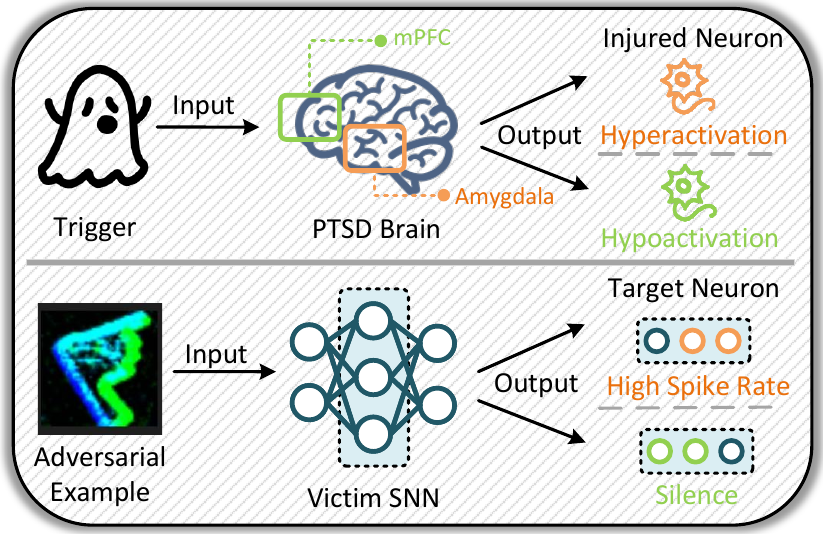}
    \caption{Motivation behind Spike-PTSD and analogies between the post-traumatic stress response of the PTSD-affected brain and AEAs against SNNs.}
    \label{fig: motivation}
\end{figure}

To address these limitations, we revisit the neurobiological foundations of SNNs. Neuroscience studies show that neurons in brains affected by Post-Traumatic Stress Disorder (PTSD) exhibit abnormal activations when exposed to specific triggers \cite{PTSD_brain, nature-ptsd}. Hyper- or hypoactivated neurons in regions such as the amygdala and the medial prefrontal cortex (mPFC) disrupt normal spike transmission, causing aberrant network responses. As shown in Fig. \ref{fig: motivation}, this phenomenon closely resembles AEAs for SNNs, where a small subset of neurons dominates spiking dynamics under adversarial perturbations. Inspired by this analogy, we propose a bio-plausible AEA framework, \textbf{Spike-PTSD}, which emulates PTSD-like neural abnormalities through controllable spike-scaling in target neurons. This mechanism bridges AEAs and biological spike dynamics for the first time, enabling higher attack success rates under temporal encoding and targeted settings. The main contributions of this work are summarized as follows:

\begin{itemize}
    \item We propose a bio-plausible AEA attack framework for SNNs by introducing abnormal phenomena of neuronal activation in PTSD-affected brains into AEAs, which is the first attempt to bridge neuroscience and adversarial machine learning to our best knowledge.
    \item We propose a universal optimization objective with spike scaling for AEAs against SNNs, which can be combined with any ANN-specific traditional optimization objectives to construct SNN-specific optimization objectives.
    \item Our method achieves over \textbf{99}\% ASR with combined optimizations under most attack settings, including targeted attacks and dynamic datasets, prompting a re-examination of SNN security designs.
\end{itemize}

\section{Preliminaries}
\subsection{Spiking Neural Networks}
\begin{figure}[t]
    \centering
    \includegraphics[width=0.99\linewidth]{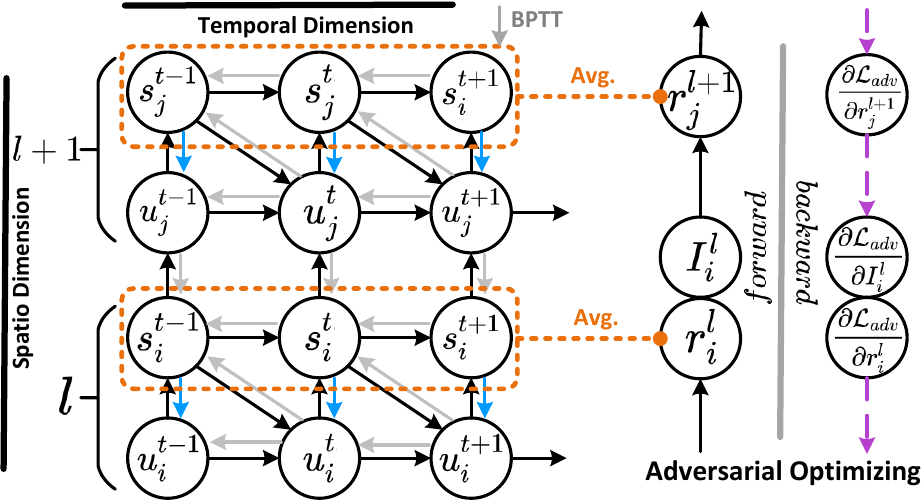}
    \caption{Forward and backward processes with surrogate gradients in direct SNN training and AEAs. $s$ and $n$ denote spike and $n$-th layer, respectively.}
    \label{fig:BPTT_RGA}
\end{figure}

SNNs primarily rely on spike neurons to capture and transmit spike information. We focus on SNNs composed of Leaky Integrate-and-Fire (LIF) neurons, as LIF neurons have been proved to exhibit high robustness in the previous work \cite{SNN_robustness_anlaysis}. The neuronal dynamics of LIF neurons can be formulated as follows:
\vspace{-0.5mm}
\begin{equation}
    \tau \frac{\mathrm{d} u}{\mathrm{d} t} =-u^t+I^t,
    \label{eq: lif_dynamics}
\end{equation}
where $u^t$ represents the neuronal membrane potential (i.e., cumulative voltage) at time $t\in\{1,\dots,T\}$. $T$ represents the total timestep, and $\tau$ denotes the decay time constant. $I_j^{t,l}=\sum_{i=1}^{\Gamma(l-1)}w_{ij}^ls_i^{t,l-1}+b_j^l$ refers to the presynaptic input $s_i^{t,l-1}$ along with synaptic weights $w_{ij}^l$ and bias $b_j^l$, where $\Gamma(l-1)$ denotes the total neuron number in the spike hidden layer $(l-1)$. SNNs process spike sequences along spatial and temporal dimensions (Fig. \ref{fig:BPTT_RGA}). During SNN inference, the mapping from membrane potential to output spikes follows the Heaviside function ($H$) as described below:
\begin{equation}
    s_i^{t,l} = H(u_i^{t,l}-u_{th}) = 
\begin{cases}
1, &  u_i^{t,l} \geq u_{th} \\
0,  &  u_i^{t,l} < u_{th},
\end{cases}
\label{eq: heaviside}
\end{equation}
which leads to a non-differentiability issue during backpropagation ($\frac{\partial s_j^{t,l+1}}{\partial u_j^{t,l+1}}$ and $\frac{\partial s_i^{t+1,l}}{\partial u_i^{t+1,l}}$):
\begin{align}
    \frac{\partial \mathcal{L} }{\partial s_i^{t,l}} =\sum_{j=1}^{\Gamma(l+1)} \frac{\partial \mathcal{L} }{\partial s_j^{t,l+1}} \frac{\partial s_j^{t,l+1}}{\partial u_j^{t,l+1}}\frac{\partial u_j^{t,l+1}}{\partial s_i^{t,l}}+ \\ \frac{\partial \mathcal{L} }{\partial s_i^{t+1,l}}  \frac{\partial s_i^{t+1,l}}{\partial u_i^{t+1,l}}\frac{\partial u_i^{t+1,l}}{\partial s_i^{t,l}} , 
    \label{eq: BPTT}
\end{align}
where $u_{th}$ represents voltage threshold for spike firing. To solve this problem, surrogate gradients are employed to ensure gradient propagation \cite{STBP}, which are highlighted by blue arrows in Fig. \ref{fig:BPTT_RGA}. Moreover, we primarily target rate-coding SNNs ($r_i^l=\frac{ {\textstyle \sum_{t=1}^{T}} s_i^{t,l}}{T}$) for two key reasons: (i) They are easier to train and the most widely used \cite{Encoding, SNN_intro}; (ii) They exhibit stronger robustness than others \cite{SNN_robustness_anlaysis}. 

\begin{figure*}[!t]
    \centering
    \includegraphics[width=0.98\linewidth]{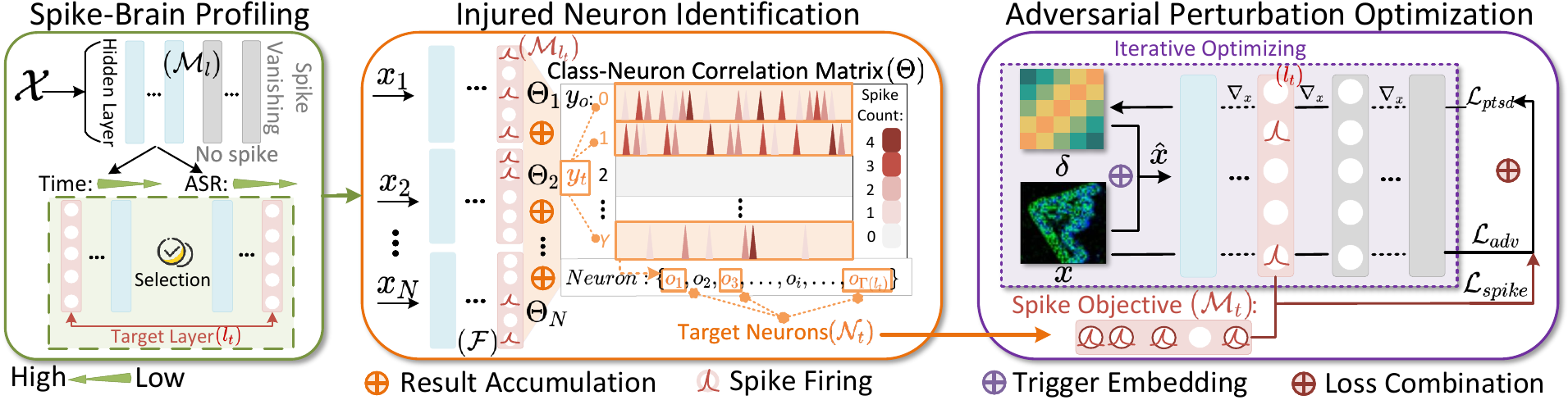}
        \caption{The overview of Spike-PTSD framework with three key modules: (i) Determine the target layer ($l_t$) during spike-brain profiling on computational time or ASR. (ii) Then, select target class ($y_t$) and neurons ($\mathcal{N}_t$) in $l_t$ according to spike-sensitivity and fired spikes under the targeted attack setting (the orange region). (iii) Finally, set the spike objective of $\mathcal{N}_t$ and update the perturbation ($\delta$) during the combined optimization process (the purple region) to generate AEs ($\hat{x}$). }
    \label{fig: method_overview}
\end{figure*}



\subsection{Adversarial Example Attacks}
AEAs can induce incorrect predictions in victim models by injecting imperceptible perturbations into input data, without modifying model parameters. The unified optimization objective can be formulated as follows:
\begin{equation}
    \arg\max_{\delta} \mathcal{L}_{adv}(\mathcal{F}(x+\delta),y)  \quad \quad s.t. \left \| \delta  \right \|_{p}<\epsilon,
    \label{eq: l_adv}
\end{equation}
where $\mathcal{L}_{adv}$ denotes the adversarial loss of AEAs, $\mathcal{F}$ represents the victim SNN, while $x$ and $y$ correspond to the input data and the target output that varies with the target setting, respectively. $\delta$ denotes the adversarial perturbation. The $L_p$-norm $\left \| \cdot  \right \|_p$ constrains the perturbation magnitude, with parameter $\epsilon$ controlling its maximum strength. FGSM \cite{FGSM} and PGD \cite{PGD} are two simple yet effective AEA methods. FGSM perturbs the original input along the direction that maximizes the model's inference loss in one iteration, as formulated below:
\begin{equation}
    \hat{x}=x+\epsilon \cdot sign(\nabla_x \mathcal{L}_{adv}(\mathcal{F}(x),y) ) ,
    \label{eq: FGSM}
\end{equation}
where $\hat{x}$ denotes the generated Adversarial Example (AE). PGD iteratively repeats this process, as defined below:
\vspace{1mm}
\begin{equation}
    \hat{x}^k=\prod _\epsilon  \left \{ x^{k-1}+\alpha \cdot sign(\nabla_x \mathcal{L}_{adv}(\mathcal{F}(x^{k-1}),y) ) \right \} ,
    \label{eq: PGD}
\end{equation}
where $k$ denotes the total iteration step, $\alpha$ controls the step size per iteration, and $\prod _\epsilon$ enforces the perturbation constraint at each step. However, as shown in the right panel of Fig. \ref{fig:BPTT_RGA}, the backpropagation during AEAs also encounters the non-differentiability problem ($\frac{\partial r^l}{\partial I^l}$), formulated as follows:
\begin{equation}
    \frac{\partial \mathcal{L}_{adv} }{\partial r^0} = \frac{\partial \mathcal{L}_{adv} }{\partial r^L} (\prod_{l=1}^{L}\frac{\partial r^l}{\partial I^l}\frac{\partial I^l}{\partial r^{l-1}}),
\end{equation}
where $L$ represents the total layer number of the victim SNN. This issue can also be resolved using the surrogate gradient for rate-coding SNNs \cite{RGA}.

\subsection{Threat Model}

\subsubsection{Attack scenario} 
Our method focuses on widely used open-source model platforms where users can upload or download models. Specifically, the attacker can be either the downloading user or the model uploader in this attack scenario.
\subsubsection{Attackers' capabilities} 
We consider targeted and untargeted attacks under white-box settings. Therefore, the attacker can access internal information of the victim model, including model structure and spike activation information. Moreover, when the attacker acts as a download user of the open-source model, they also gain access to the test dataset associated with the victim model, which is used to analyze the internal spiking information of the victim model and generate AEs.

\section{Methodology}

In this section, we present our bio-plausible AEA framework (Spike-PTSD), which simulates the post-traumatic stress responses in the PTSD-affected brain with corresponding spike scalings through three key modules: (i) spike-brain profiling, (ii) injured neuron identification, and (iii) adversarial perturbation optimization. Fig. \ref{fig: method_overview} illustrates the complete framework of Spike-PTSD.

\subsection{Spike-Brain Profiling (SBP)}
In this module, we identify target layers ($l_t$) by analyzing spike activation maps ($\mathcal{M}_l$) in the victim model ($\mathcal{F}$), mirroring the functional roles of the amygdala (or mPFC) in the PTSD-affected brain. First, for each spike hidden layer $l$ in $\mathcal{F}$, we obtain the corresponding $\mathcal{M}_l$ (Eq.\ref{eq: spike_map}) during the inference process to determine whether spike vanishing occurs. 
\begin{equation}
    \{\mathcal{M}_l\}=\{\mathcal{M}_1,\mathcal{M}_2,\dots ,\mathcal{M}_{L-d}\} = hook_{1:L-d}\{\mathcal{F}(\mathcal{X})\},
    \label{eq: spike_map}
\end{equation}
where $d$ denotes the number of decision layers. $\mathcal{X}=\{x_i,y_i\}_{i=1}^N$ denotes the test dataset, where $N$ is the total number of test data samples, and $y_i$ represents the origin label of $x_i$. $hook\{\cdot\}$ represents the spike extraction function. If spike vanishing occurs, spike hidden layers with vanished spikes are removed, and the remaining spike hidden layers are considered as target candidates for $l_t$. Then, $l_t$ will serve as the search space for the target neuron $\mathcal{N}_t$ in the next module.

\subsection{Injured Neuron Identification (INI)}
In the second module, we identify target neurons $\mathcal{N}_t$ by mirroring hyper- or hypoactivated neurons in the PTSD-affected brain under post-traumatic stress responses. We accordingly develop dual localization schemes for targeted and untargeted attacks. 
\textbf{For targeted attacks}, we identify target neurons $\mathcal{N}_t$ in $l_t$, analogous to amygdala hyperactivation in the PTSD-affected brain that induces specific misbehaviour. As shown in the orange region in Fig. \ref{fig: method_overview}, the procedure comprises three steps: i) Each test sample $x_i$ is sequentially fed into the victim model ($\mathcal{F}$) to obtain its output class $y_o$ and the corresponding spike activation map $\mathcal{M}_{l_t}$ at the target layer to construct class-neuron correlation matrix $\Theta_i \in \mathbb{R} ^{Y \times \Gamma(l_t)}$, where $Y$ denotes the total number of classes. Specifically, if $\mathcal{F}(x)=y_o$ and $r_i=\mathcal{M}_{l_t}[o_i]\ne0$ during inference, where $r_i$ represents the spike rate of $o_i$ and $o_i$ denotes neuron index in $l_t$. We set $\Theta_i[y_o][o_i]=1$; otherwise, $\Theta_i[y_o][o_i]=0$. Then, $\Theta_i$ of all test samples is aggregated to construct the final correlation matrix $\Theta=\Theta_1\oplus \Theta_2\oplus \cdots \oplus \Theta_N$. (ii) Select classes reliant on spike activation for inference as the target class $y_t$, i.e., $y_t =y_o$ if $\exists o_i, \Theta[y_o][o_i]\ne 0$. The intuition behind this is to analogize $y_t$ to the specific behaviours resulting from neuron hyperactivation in the Amygdala during stress responses. (iii) Extract critical neurons that significantly contribute to the inference of $y_t$ as $\mathcal{N}_t$, i.e., $\mathcal{N}_t=\{o_i|\Theta[y_t][o_i]\ne 0 \} $. 

\begin{figure}[!t]
    \centering
    \includegraphics[width=0.99\linewidth]{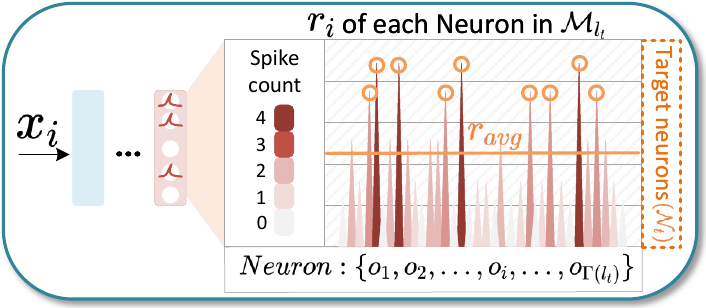}
    \caption{Specific process of INI for untargeted attacks.}
    \label{fig: untar_INI}
\end{figure}

\textbf{For untargeted attacks}, we identify neurons in $l_t$ analogous to hypoactivated neurons in the mPFC that cause regulatory dysfunction as $\mathcal{N}_t$. As shown in Fig. \ref{fig: untar_INI}, our selection strategy is to inhibit high-spike-rate neurons.
Generally, the average pooling layer is added after the spike hidden layer in SNNs to downsample the spike output of the spike layer, and the specific output of the averaging pooling layer can be expressed as follows:
\begin{equation}
    Avg_l=\frac{1}{\left | \Gamma _{(l)} \right | } \sum_{o_i\in\Gamma_{(l)}}\mathcal{M}_l[o_i]=\frac{1}{\left | \Gamma _{(l)} \right | } \sum_{i=1}^{\Gamma(l)} r_i,
\end{equation}
Thus, the average spike rate $r_{avg}$ of $l_t$ can be calculated as:  $r_{avg}=\frac{1}{\Gamma(l_t)} \sum_{i=1}^{\Gamma(l_t)} r_i$. Under untagreted attacks, we aim to suppress highly contributing neurons with high spike rates ($r_i>r_{avg}$) as $\mathcal{N}_t$ in $l_t$ to keep silent. The underlying reason is that: the ideal output spike map $\mathcal{M}_{l_t}$ can be formulated as:
\begin{equation}
\hat{\mathcal{M}}_{l_t}[o_i]=
\begin{cases}
0, &  \mathrm{if}  \quad r_i > r_{avg} \\
\mathcal{M}_{l_t}[o_i],  & \mathrm{otherwise} \\
\end{cases}
,
\end{equation}
Then, the adversarial $\hat{Avg}_{l_t}$ after the spike suppression in $\hat{\mathcal{M}}_{l_t}$ can be expressed as:
\begin{equation}
    \hat{Avg}_{l_t}= \frac{1}{\left | \Gamma(l_t) \right |}\sum_{o_i\in\Gamma(l_t),r_i\le r_{avg}}\mathcal{M}_{l_t}[o_i],
\end{equation}
Thus, the spike variation ($\Delta Avg_{l_t}$) can be formulated as:
\begin{equation}
    \Delta Avg_{l_t}=Avg_{l_t}-\hat{Avg}_{l_t} =\frac{1}{\left | \Gamma(l_t) \right | } \sum_{o_i\in \mathcal{H}_{l_t}}\mathcal{M}_{l_t}[o_i],  
    \label{eq: avg_variation}
\end{equation}
where $\mathcal{H}_{l_t}$ denotes the set of high-spike-rate neurons in $\mathcal{M}_{l_t}$. Then, if the significance condition in Eq. \ref{eq: significant_condition} is satisfied, it can be verified that the spike contributions of the neurons in $\mathcal{H}_{l_t}$ significantly dominate the model inference.
\vspace{1mm}
\begin{equation}
    \mathcal{D}_r=\frac{ {\textstyle \sum_{k\in\mathcal{H}_{l_t}}}r_k}{ {\textstyle \sum_{k\in\Gamma(l_t)}} r_k} >\frac{\left | \mathcal{H}_{l_t}  \right | }{\left | \Gamma(l_t) \right | } = \mathcal{D}_{l_t}.
    \label{eq: significant_condition}
\end{equation}
\vspace{3mm}
Due to the sparse spiking activations of SNNs, this significance condition generally holds for most network architectures. Moreover, we experimentally verified the significance condition for all models involved in this work (Sec. \ref{sec: condition_analysis}). 

\subsection{Adversarial Perturbation Optimization (APO)}

In the last module of Spike-PTSD, we introduce spike scaling as an additional optimization objective ($\mathcal{L}_{spike}$) for both targeted and untargeted attacks with the original adversarial optimization objective (Eq. \ref{eq: l_adv}), which can be further formulated as:
\vspace{1mm}
\begin{equation}
\mathcal{L}_{adv}=
\begin{cases}
\sum_{i=1}^{N}\mathcal{L}_{ce}(\mathcal{F}(x_i+\delta), y_t) ,& targeted \\
\sum_{i=1}^{N}\mathcal{L}_{ce}(\mathcal{F}(x_i+\delta),y_i) , & untargeted,
\end{cases}
\label{eq: soft_reset}
\end{equation}
Then, $\mathcal{L}_{spike}$ for spike scaling under targeted and untargeted attacks can be uniformly expressed as:
\begin{equation}
    \mathcal{L}_{spike} = \mathcal{L}_{mse}(\mathcal{M}_{l_t}, \mathcal{M}_t); \quad \mathcal{M}_t= \{r_i|\mathcal{M}_{l_t} [\mathcal{N}_t]= r_t\},
    \label{eq: spike_objective}
\end{equation}
where $\mathcal{M}_t$ represents the target spike objective and $r_t$ denotes the expected spike rate of $\mathcal{N}_t$.
\textbf{For targeted attacks}, we set spike amplification as the final optimization objective for $\mathcal{N}_t$, i.e., $r_t=1$. \textbf{For untargeted attacks}, we expect that $\mathcal{N}_t$ can remain silent (spike suppression) during inference on each AE, i.e., $r_t=0$. Considering that both $\mathcal{L}_{adv}$ (targeted attack) and $\mathcal{L}_{spike}$ are loss minimization processes, while the overall optimization goal of AEA (Eq. \ref{eq: l_adv}) is a loss maximization process through gradient ascent. Therefore, a negative sign needs to be added to the corresponding $\mathcal{L}_{adv}$ and $\mathcal{L}_{spike}$ in the combined optimization objective $\mathcal{L}_{ptsd}$. The final $\mathcal{L}_{ptsd}$ for bio-plausible Spike-PTSD is as follows:
\begin{equation}
\mathcal{L}_{ptsd}=
\begin{cases}
-\gamma \mathcal{L}_{adv}-\beta \mathcal{L}_{spike},  &  targeted \\
\lambda \mathcal{L}_{adv}-\mu \mathcal{L}_{spike},    & untargeted,
\end{cases}
\label{eq: com_opti_objectives}
\end{equation}
where $\gamma$, $\beta$, $\lambda$, and $\mu$ are coefficients to trade off the two optimization objective terms. Finally, the AE ($\hat{x}$) is generated through $k$ iterations during the perturbation optimization according to Eq. \ref{eq: FGSM} (or \ref{eq: PGD}). 


\section{Experiments}
This section outlines the basic experimental setup and the main experimental results. All experiments were conducted on a server with eight NVIDIA RTX 3090 GPUs.

\subsection{Experimental Setups}
\subsubsection{Datasets and Models}We evaluate Spike-PTSD and state-of-the-art (SOTA) works on six datasets and three encoding types ($E$): \textbf{(i) Traditional:} CIFAR10~\cite{CIFAR10_CIFAR100}, CIFAR100~\cite{CIFAR10_CIFAR100}, and SVHN~\cite{SVHN} with direct ($E_1$) and Poisson ($E_2$) encodings; \textbf{(ii) Neuromorphic:} CIFAR10-DVS (CIFARDVS)~\cite{CIFAR10-DVS}, N-MNIST~\cite{N-MNIST}, and DVS128Gesture (Gesture)~\cite{DVS128Gesture} with frame ($E_3$) encoding. Moreover, the experiments are conducted on four spike-based model structures: VGG16, ResNet18 (Res18), VGGDVS, and ResNet19DVS (ResDVS). 

\subsubsection{Metric}
Attack Success Rate (ASR) is a crucial metric for evaluating the attack performance of AEA. Targeted and untargeted AEAs are evaluated based on the accuracy of the victim model for correct and wrong classification of $\hat{x}$, respectively, formulated as follows:
\begin{equation}
\begin{cases}
ASR_{tar}= \frac{1}{N} \sum_{i=1}^{N} \mathbb{I} (\mathcal{F}(\hat{x}_i) = y_t )\times 100\% \\
ASR_{untar}= \frac{1}{N} \sum_{i=1}^{N} \mathbb{I} (\mathcal{F}(\hat{x}_i)\ne y_i )\times 100\%
\end{cases}
.
\label{eq: metrics_asr}
\end{equation}

\subsubsection{Hyperparameter Settings}
For SNN models, $\tau=0.2$, and $u_{th}=1.0$. During perturbation optimization, $\left \| \cdot \right \| _p =\left \| \cdot \right \| _\infty $ and $\epsilon = 2/255$ are used for FGSM and PGD by default. Moreover, we set $\gamma=\beta=\lambda=\mu=1$ to balance the loss terms in the combined optimization objective (Eq. \ref{eq: com_opti_objectives}) by default. 

\begin{table*}[]
\centering
\caption{Attack Performance Comparison with SOTA works on static and dynamic datasets.}
\vspace{-4mm}
\scalebox{0.87}{
\begin{tabular}{ccc|rrrrr|rrrr}
\toprule
                                    &                                                 &                              & \multicolumn{5}{c|}{\textbf{Targeted ASR ($\pm SE$)}}                                                                                                                                & \multicolumn{4}{c}{\textbf{Untargeted ASR ($\pm SE$)}}                                                                                  \\ \cline{4-12} 
\multirow{-2}{*}{\textbf{Dataset}}  & \multirow{-2}{*}{\textbf{Model}}                & \multirow{-2}{*}{\textbf{$E$}} & \multicolumn{1}{l|}{\textbf{$y_t$}}                & \textbf{RGA \cite{RGA}}                 & \textbf{HART \cite{HART}}                 & \textbf{PDSG \cite{PDSG}}             & \textbf{Ours ($\uparrow$)}                 & \textbf{RGA \cite{RGA}}                  & \textbf{HART \cite{HART}}                 & \textbf{PDSG \cite{PDSG}}                 & \textbf{Ours ($\uparrow$)}                 \\ \hline
                                    &                         & $E_1$   & \multicolumn{1}{r|}{6}  & 0.5 ($\pm0.5$)   & \underline{41.65 ($\pm 0.2$)} & 0 ($\pm 0$)& \textbf{99.82 ($\pm$ 0.1)}& 74.47 ($\pm 0.6$)& \underline{88.47 ($\pm 0.5$)}& 62.87 ($\pm 0.3$)& \textbf{99.13 ($\pm$ 0.7)} \\
                                    & \multirow{-2}{*}{Res18} & $E_2$   & \multicolumn{1}{r|}{9}  & 0 ($\pm 0$)   & \underline{9.43 ($\pm0.5$)}  & 0 ($\pm0$)& \textbf{90.66 ($\pm$0.7)} & 18.8 ($\pm0.3$) & \underline{30.07 ($\pm$0.5)} & 13.41 ($\pm0.4$)& \textbf{56.23 ($\pm$0.7)}   \\
                                    &                                                 & $E_1$                           & \multicolumn{1}{r|}{8}                          & 1.67 ($\pm0.3$)                        & \underline{65.7 ($\pm0.5$)}                          & 0 ($\pm0$)                        & \textbf{97.49 ($\pm$0.5)}                        & 87.19 ($\pm0.3$)                        & \underline{93.09 ($\pm0.4$)}                        & 84.19 ($\pm0.3$)                        & \textbf{99.28 ($\pm$0.7)}                        \\
\multirow{-4}{*}{\textbf{CIFAR10}}  & \multirow{-2}{*}{VGG16}                         & $E_2$                           & \multicolumn{1}{r|}{1}                          & 1.49 ($\pm0.3$)                        & \underline{11.12 ($\pm0.6$)}                        & 0 ($\pm0$)                        & \textbf{77.06 ($\pm$0.8)} & 17.12 ($\pm0.5$)                         & \underline{28.99 ($\pm0.2$)}                        & 10.71 ($\pm0.3$)                       & \textbf{53.64 ($\pm$0.8)} \\ \hline
                                    &                         & $E_1$   & \multicolumn{1}{r|}{6}  & 0 ($\pm0$)   & \underline{32.59 ($\pm0.4$)} & 0 ($\pm0$)& \textbf{99.8 ($\pm$0.1)} & 58.46 ($\pm0.5$)    & \underline{86.57 ($\pm0.3$)} & 42.41 ($\pm0.5$) & \textbf{98.75 ($\pm$0.7)}\\
                                    & \multirow{-2}{*}{Res18} & $E_2$   & \multicolumn{1}{r|}{0}  & 0 ($\pm0$)   & \underline{8.32} ($\pm0.6$) & 0 ($\pm0$)& \textbf{95.16 ($\pm$0.7)} & 17.89 ($\pm0.2$)    & \underline{25.25 ($\pm0.3$)} & 10.53 ($\pm0.3$)& \textbf{47.52 ($\pm$0.7)}    \\
                                    &                                                 & $E_1$                           & \multicolumn{1}{r|}{2}                          & 1.01 ($\pm0.6$)                        & \underline{38.08 ($\pm0.5$)}                       & 0.83 ($\pm0.4$)                     & \textbf{97.82 ($\pm$0.2)}                        & 55.25 ($\pm0.7$)                        & \underline{63.36 ($\pm0.5$)}                        & 45.47 ($\pm0.5$)                        & \textbf{99.86 ($\pm$0.1)}                        \\
\multirow{-4}{*}{\textbf{SVHN}}     & \multirow{-2}{*}{VGG16}                         & $E_2$                           & \multicolumn{1}{r|}{2}                          & 0.26 ($\pm0.1$)                        & \underline{9.28 ($\pm0.5$)}                         & 0 ($\pm0$)                         & \textbf{91.67 ($\pm$0.7)} & 16.34 ($\pm0.6$)                         & \underline{27.02 ($\pm0.7$)}                        & 10.14 ($\pm0.6$)                        & \textbf{42.28 ($\pm$0.7)} \\ \hline
                                    & Res18                   & $E_1$   & \multicolumn{1}{r|}{15} & 2.31 ($\pm0.5$)& \underline{43.12} ($\pm0.7$) & 0 ($\pm0$) & \textbf{99.87 ($\pm$0.1)} & \underline{93.21 ($\pm0.7$)}    & 86.57 ($\pm0.5$) & 81.62 ($\pm0.7$)& \textbf{99.42} ($\pm$0.5)    \\
\multirow{-2}{*}{\textbf{CIFAR100}} & VGG16                                           & $E_1$                           & \multicolumn{1}{r|}{90}                         & 2.11 ($\pm0.7$)                        & \underline{63.8 ($\pm0.5$)}                         & 0 ($\pm0$)                        & \textbf{98.27 ($\pm$0.7)}                       & 91.15 ($\pm0.3$)                        & \underline{94.28 ($\pm0.5$)}                        & 90.08 ($\pm0.4$)                        & \textbf{99.46 ($\pm0.5$)}                           \\ \hline
                                    & VGGDVS                                          & $E_3$                           & \multicolumn{1}{r|}{1}                          & 16.04 ($\pm0.2$)                        & \underline{32.73 ($\pm0.5$)}                        & 0 ($\pm0$)                        & \textbf{99.39 ($\pm$0.6)}                         & 72.14 ($\pm0.7$)                        & \underline{73.68 ($\pm0.5$)}                        & 53.38 ($\pm0.8$)                        & \textbf{99.23 ($\pm$0.7)}                        \\
\multirow{-2}{*}{\textbf{CIFARDVS}} & ResDVS                                          & $E_3$                           & \multicolumn{1}{r|}{8}                          & 15.57 ($\pm0.4$)                       & \underline{45.16 ($\pm0.7$)}                        & 0 ($\pm0$)                         & \textbf{99.57 ($\pm$0.4)}                        & 84.14 ($\pm0.6$)                        & \underline{84.19 ($\pm0.5$)}                        & 51.3 ($\pm0.5$)                         & \textbf{99.14 ($\pm$0.8)}                        \\ \hline
                                    & VGGDVS                                          & $E_3$                           & \multicolumn{1}{r|}{0}                          & 0.38 ($\pm0.2$)                        & \underline{0.92 ($\pm0.5$)}                         & 0 ($\pm0$)                        & \textbf{99.6 ($\pm$0.3)}                          & 1.7 ($\pm0.5$)                           & 1.64 ($\pm0.5$)                          & \underline{2.28 ($\pm0.7$)}                          & \textbf{99.59 ($\pm$0.4)}                        \\
\multirow{-2}{*}{\textbf{N-MNIST}}  & ResDVS                                          & $E_3$                           & \multicolumn{1}{r|}{0}                          & 0.54 ($\pm0.3$)                        & \underline{1.36 ($\pm0.2$)}                          & 0.91 ($\pm0.2$)                     & \textbf{92.13 ($\pm$0.5)}                        & 2.02 ($\pm0.2$)                         & \underline{2.12 ($\pm0.2$)}                         & 1.92 ($\pm0.3$)                         & \textbf{72.34 ($\pm$0.5)}                        \\ \hline
                                    & VGGDVS                                          & $E_3$                           & \multicolumn{1}{r|}{9}                          & 63.24 ($\pm0.5$)                       & \underline{69.56 ($\pm0.8$)}                        & 0 ($\pm0$)                        & \textbf{99.46 ($\pm$0.5)}                        & 82.21 ($\pm0.5$)                        & \underline{82.41 ($\pm0.6$)}                        & 77.47 ($\pm0.3$)                        & \textbf{99.45 
                                    ($\pm$0.5)}\\
\multirow{-2}{*}{\textbf{Gesture}}  & ResDVS                                          & $E_3$                           & \multicolumn{1}{r|}{10}                         & 0.54 ($\pm0.2$)                        & \underline{55.69 ($\pm0.5$)}                         & 0 ($\pm0$)                        & \textbf{92.13 ($\pm$0.7)}                        & \underline{89.43 ($\pm0.5$)}                        & 89.02 ($\pm0.6$)                        & 84.15 ($\pm0.5$)                        & \textbf{99.44 ($\pm$0.5)}                          \\ \bottomrule
\multicolumn{12}{l}{$\uparrow$ means the higher the better. The best results are in bold, and the second-best results are underlined.}
\end{tabular}
}
\label{tab: overall_performance}
\end{table*}

\begin{figure*}[!t]
    \footnotesize
    \centering
    \subfigure[Spike Layers]{\label{fig: ablation_layer}    {\includegraphics[width=0.62\textwidth]{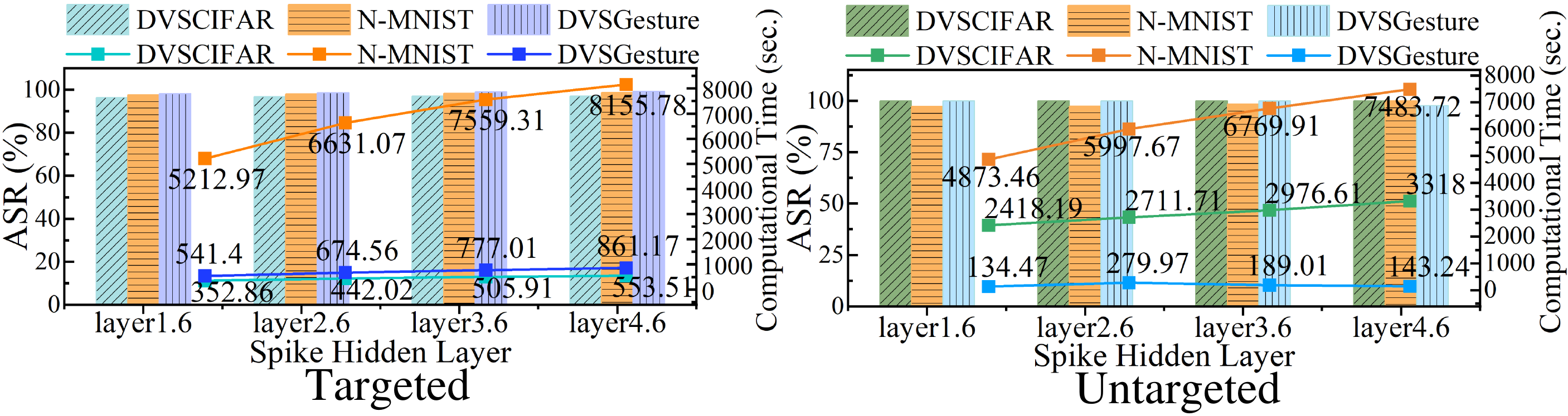}}}\hfill
    \subfigure[Hyperparameters]{\label{fig: ablation_param}
    {\includegraphics[width=0.365\textwidth]{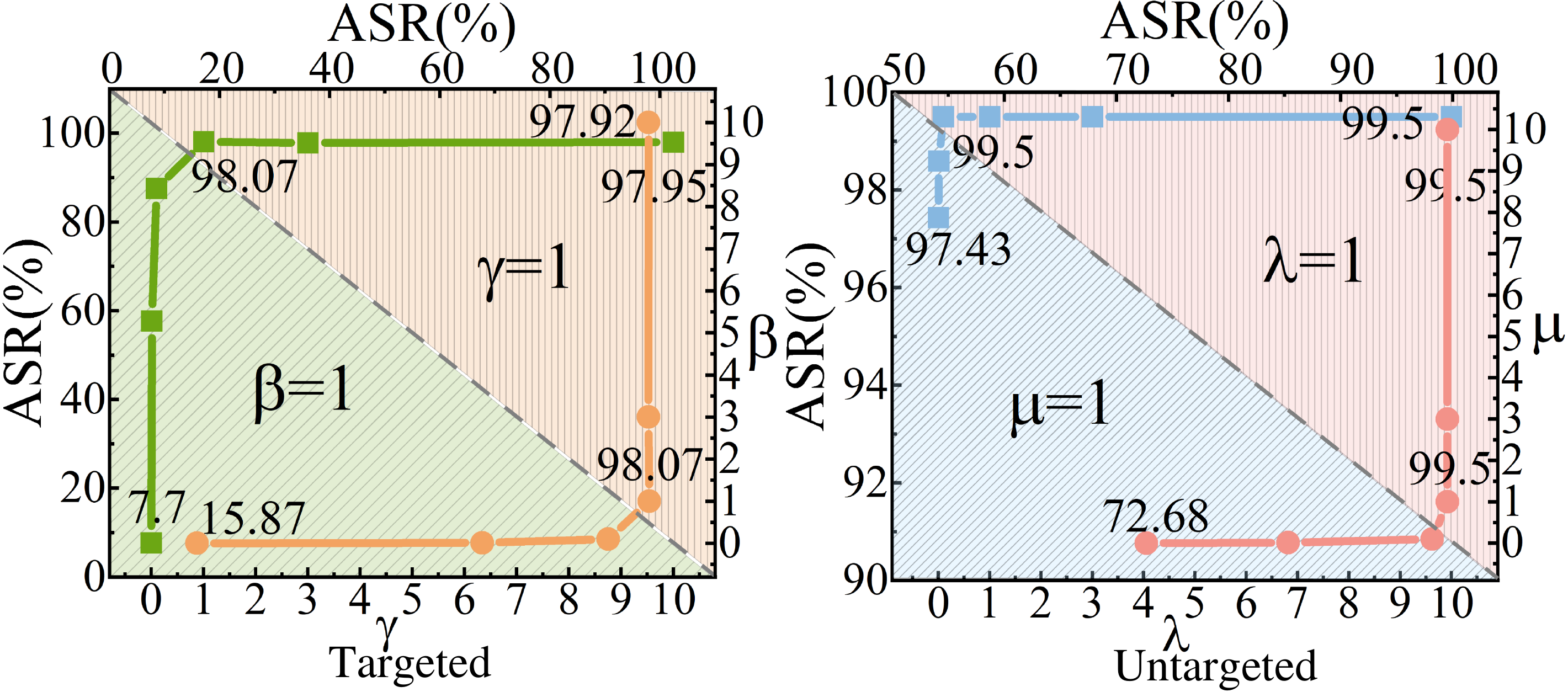}}}\hfill
    \caption{Ablation studies on different target spike layers in VGGDVS (a) and hyperparameters (b).}
    \label{fig: ablation_SBP}
\end{figure*}

\subsection{Comparison with State-of-the-art Works}
We comprehensively compared the attack performance of Spike-PTSD with SOTA works, including RGA \cite{RGA}, HART \cite{HART}, and PDSG \cite{PDSG}. The results are presented in TABLE \ref{tab: overall_performance}, which records the average ASR across 20 independent runs, with Standard Error ($\pm SE$) in the outcomes. For different datasets and model architectures, we employ the PGD method for all compared approaches, using the default parameters $k = 15$ and $\epsilon = 2/255$. The data shows Spike-PTSD consistently achieves the highest ASR across all three encoding schemes, regardless of whether the dataset is static or dynamic. Notably, while other SOTA methods yield nearly zero ASR under targeted attack settings, Spike-PTSD still attains over 99\% ASR in most experimental configurations. 

\subsection{Ablation studies}

\subsubsection{Ablation Study on Spike Hidden Layers}

To validate the effectiveness of target layer candidates $l_t$ in SBP, we conducted ablation experiments under different spike hidden layers. The specific attack performance and computational overhead under targeted and untargeted attacks on VGGDVS are documented in Fig. \ref{fig: ablation_layer}. We observe that for targeted attacks, an increase in target layer depth is accompanied by a slight rise in ASR and computational time, which is not significant under untargeted attacks. However, the situation on DVS128Gestue is an exception. We believe this is due to the enormous disparity in the number of target neurons across spike layers, as layer2.6 contains 220,000 target neurons while all other layers have fewer than 50,000. Therefore, shallower spike layers incur lower temporal overhead, whereas deeper spike hidden layers achieve higher ASR. 




\subsubsection{Ablation Study on Hyperparameters}
Fig. \ref{fig: ablation_param} reports the ASR variations under both targeted and untargeted attacks with different hyperparameter values. Note that $\gamma=\lambda=0$ removes the adversarial term, while $\beta=\mu=0$ removes the spike scaling term. As shown in Fig. \ref{fig: ablation_layer}, when $\beta$ and $\gamma$ are properly combined, they can significantly boost ASR from 7.7\% to 98.07\% (or from 15.87\% to 98.07\%). Moreover, ASR is maximized when $\gamma=1$ and $\beta=1$, whereas further increasing either term may cause a slight degradation. In contrast, the spike scaling term can almost replace the adversarial term under untargeted attacks, since Spike-PTSD still achieves an ASR of 97.43\% even when $\lambda=0$. 

\begin{figure}[!t]
    \centering
    \includegraphics[width=0.98\linewidth]{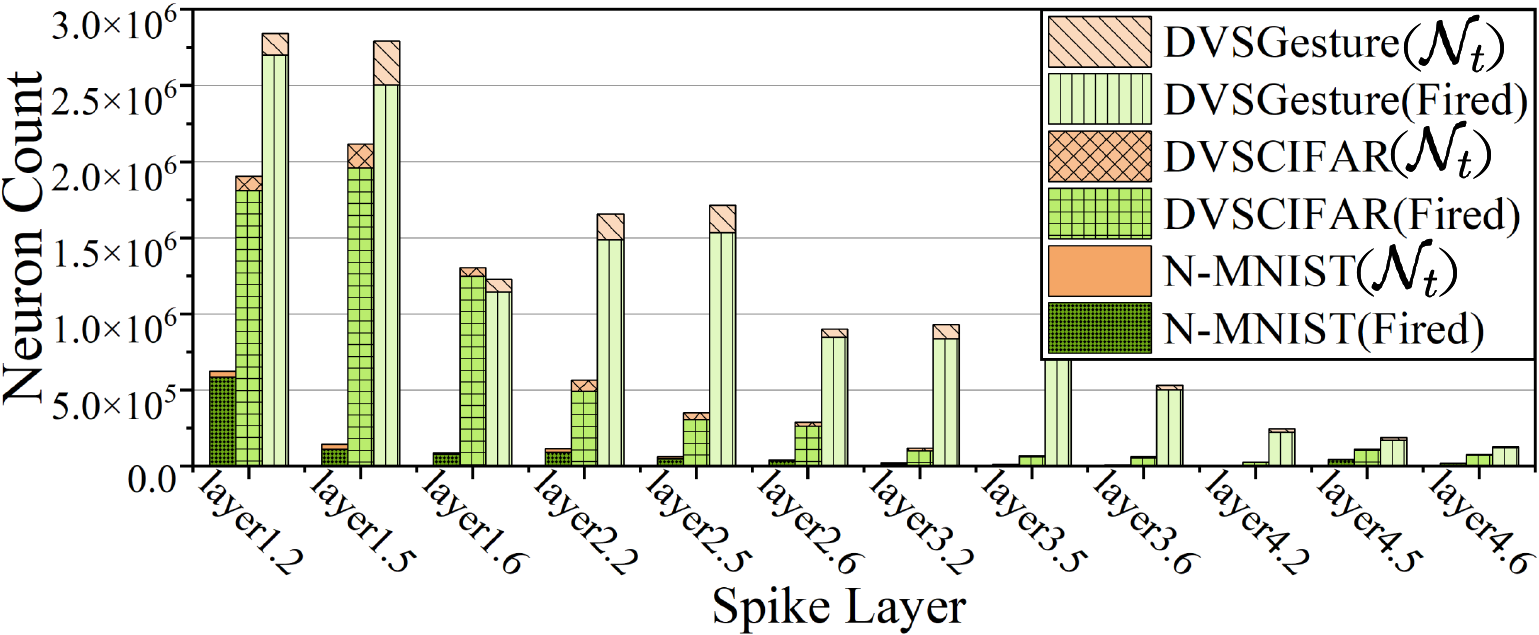}
    \caption{The number of fired neurons in the spike layers of VGGDVS and the number of $\mathcal{N}_t$ under targeted attacks.}
    \label{fig: fired_target_neuron}
\end{figure}

\subsection{Effectiveness of Spike Scaling}

\begin{figure}[!t]
    \centering
    \includegraphics[width=0.998\linewidth]{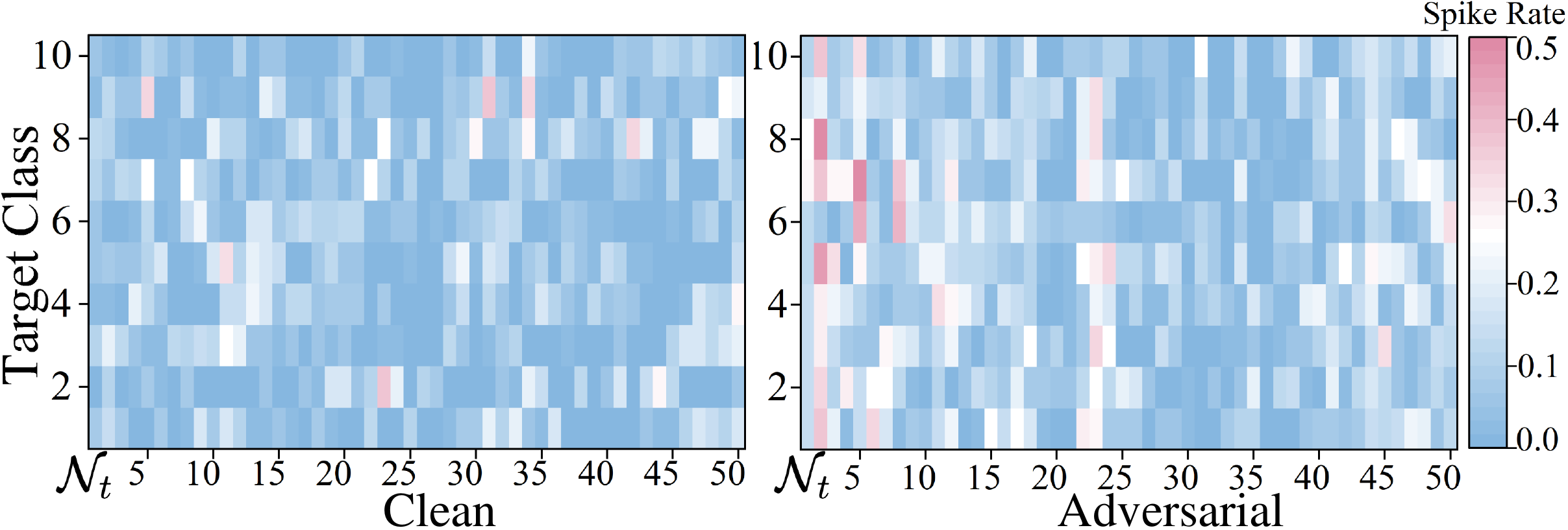}
    \caption{Spike rate changes under targeted attack ($y_t=1$) with spike amplification objectives on CIFAR10-DVS (VGGDVS). The data is intercepted from the first 50 neurons in $\mathcal{N}_t$.}
    \label{fig: tar_spike}
\end{figure}

\begin{figure}[!t]
    \centering
    \includegraphics[width=0.99\linewidth]{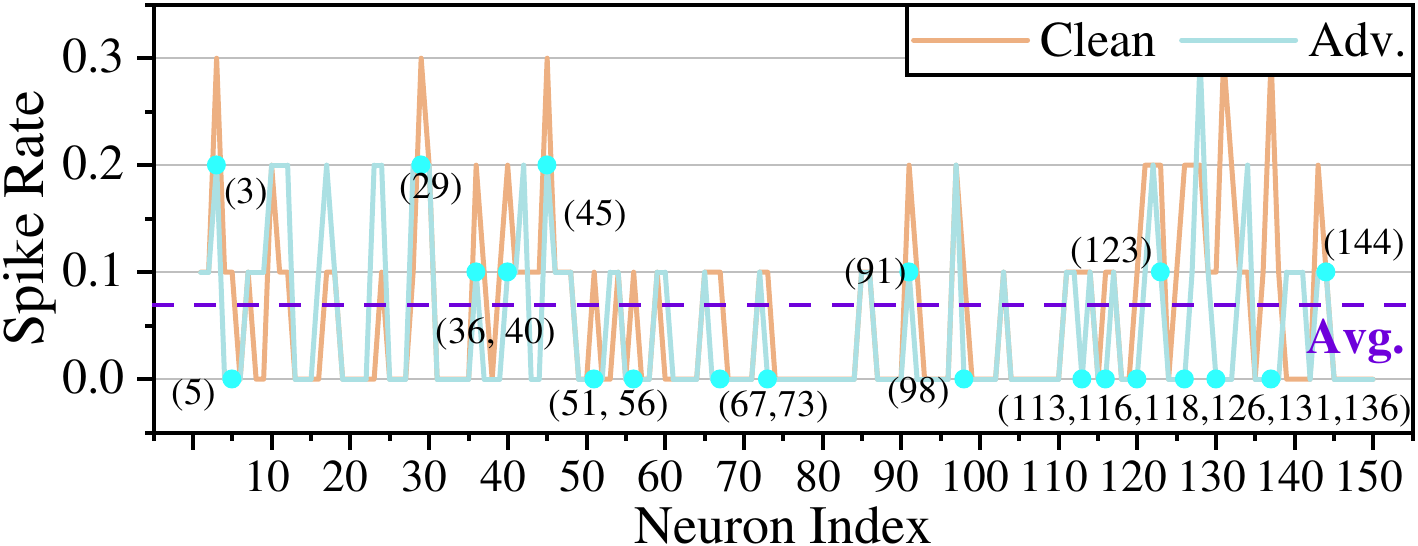}
    \vspace{-2mm}
    \caption{Spike rate changes under untargeted Spike-PTSD with spike suppression objectives on CIFAR10-DVS (VGGDVS). Neurons with $r_i$ above the \textbf{Avg.} on the clean line are $\mathcal{N}_t$.}
    \label{fig:untar_spike}
\end{figure}

Targeted and untargeted attacks adopt distinct spike scaling strategies, namely spike amplification for targeted attacks and suppression for untargeted attacks. As shown in Fig. \ref{fig: fired_target_neuron}, $\mathcal{N}_t$ is identified after the SBP and INI procedures, which is only a small fraction of the activated neurons in $l_t$. 
\textbf{For targeted attacks}, the changes in spike rates under spike amplification effects are recorded in Fig. \ref{fig: tar_spike}. The data shows that the AE generated by targeted Spike-PTSD significantly enhances the $r_i$ of $\mathcal{N}_t$, especially for $\mathcal{N}_t[0-27]$ and $\mathcal{N}_t[35-50]$. \textbf{For untagrted attacks}, the spike rate variations under spike suppression effects are recorded in Fig. \ref{fig:untar_spike}. As expected, the AE generated under untargeted settings caused the $r_i$ of the most neurons in $\mathcal{N}_t$ to plummet or even fall silent.




\subsection{Discussion}

\subsubsection{Significance Condition Analysis}
\label{sec: condition_analysis}
The experimental results of the significance condition under most of the attack settings are recorded in Table \ref{tab: feasiblity_untar_INI}. It can be seen that $\mathcal{D}_r$ is significantly higher than $\mathcal{D}_{l_t}$ in all settings, and thus inhibiting high-spike-rate neurons can lead to significant loss of important spike information and model performance.

\subsubsection{Attack Robustness}
\begin{figure}[!t]
    \footnotesize
    \centering
    \subfigure[Different Target Classes]{\label{fig:diff_TC}
    {\includegraphics[width=0.23\textwidth]{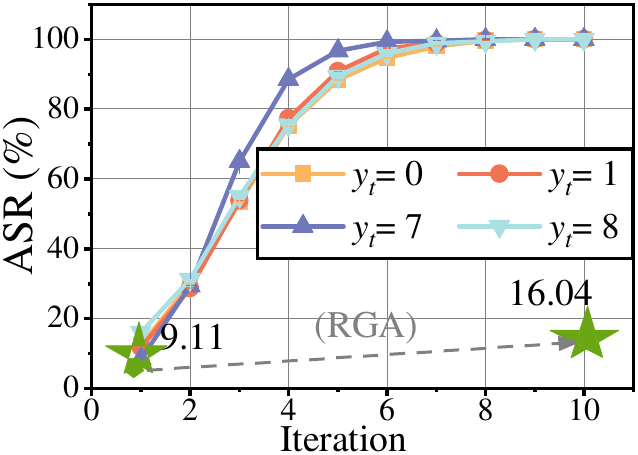}}}\hfill
    \subfigure[Different Timesteps]{\label{fig:diff_Tiemstep}
    {\includegraphics[width=0.232\textwidth]{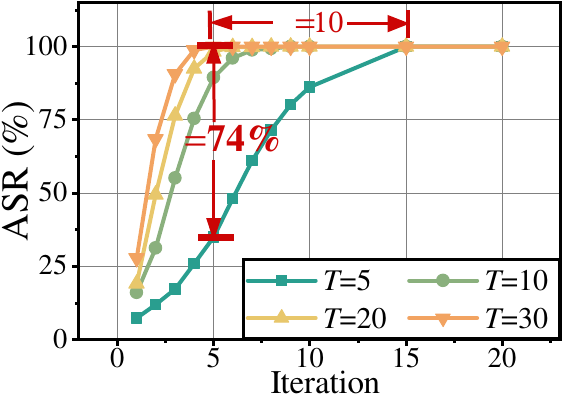}}}\hfill
    \vspace{-3mm}
    \caption{Backdoor attack performance under N-MNIST , CIFAR10-DVS, and DVS128Gesture on VGGDVS with different target classes and timesteps.}
    \label{fig:LR_B-Neuromorphic-dataset}
\end{figure}

This section analyzes the attack robustness of Spike-PTSD and the inherent robustness of SNN under different target classes $y_t$ and timestep $T$. The specific experimental results are recorded in Fig. \ref{fig:diff_TC} and  \ref{fig:diff_Tiemstep}. It shows that the attack performance (ASR) of Spike-PTSD under different $y_t$ all significantly improved as the optimization iteration increased (Fig. \ref{fig:diff_TC}). In contrast, the baseline method (RGA \cite{RGA}) only shows a limited ASR improvement (9.11\%$\to$16.04\%). Moreover, as shown in Fig. \ref{fig:diff_Tiemstep}, the ASRs of Spike-PTSD for victim models under different $T$ can generally stabilize at $\approx$100\% as the optimization iteration increases. Therefore, we believe Spike-PTSD has high attack robustness under most attack settings. For the inherent robustness of SNNs, the previous study argued that SNNs exhibit higher robustness than the ANN under equal strength AEAs, since they process discrete inputs and use nonlinear activation \cite{SNN_robustness_anlaysis}. However, Spike-PTSD can steadily increase the ASR to nearly 100\%, which completely breaks through the inherent robustness of SNNs. Additionally, we observe that SNNs with smaller $T$ achieve higher robustness, as more iterations are required for Spike-PTSD to achieve $\approx$100\% ASR under the same experimental setup (Fig. \ref{fig:diff_Tiemstep}).


\begin{table}[!t]
\centering
\caption{Significant contributions of high-spike-rate neurons.}
\vspace{-2mm}
\begin{tabular}{cccc}
\toprule
\textbf{Dataset} & \textbf{Model} & \textbf{$\mathcal{D}_r$} & \textbf{$\mathcal{D}_{l_t}$} \\ \midrule[0.7pt]
CIFAR10-DVS         & VGGDVS         & $\approx$ 1                       & 0.4601                   \\
                 & ResNet19DVS       & $\approx$ 1                       & 0.1589                   \\
DVS128Gesture       & VGGDVS         & 0.9969                   & 0.3148                   \\
                 & ResNet19DVS       & $\approx$ 1                       & 0.2317                   \\
N-MNIST           & VGGDVS         & $\approx$ 1                       & 0.2714                   \\
                 & Res19NetDVS       & $\approx$ 1                       & 0.0877                   \\
CIFAR10          & VGG16          & $\approx$ 1                       & 0.1881                   \\
SVHN             & VGG16          & 0.9064                   & 0.3805                   \\ \bottomrule
\end{tabular}
\label{tab: feasiblity_untar_INI}
\end{table}




\subsubsection{System and Hardware Implications}

Beyond algorithmic insights, Spike-PTSD also reveals system-level vulnerabilities that are highly relevant to neuromorphic hardware security. In hardware implementations of SNNs, such as Loihi \cite{Loihi} or TrueNorth \cite{TrueNorth}, spike scaling as modeled in Spike-PTSD can be physically manifested as variations in firing thresholds, membrane voltage perturbations, or spike-timing delays caused by hardware noise or an intentional signal manipulation \cite{hardware_trojan, ETBT}. These effects may emerge either from hardware Trojans embedded in synaptic circuits or from maliciously crafted event streams injected through neuromorphic sensors \cite{hardware_trojan_survey}. The observation that hyper- and hypoactivated neurons can globally destabilize the spike dynamics of SNNs implies that even minor analog perturbations or timing shifts at the circuit level could lead to functional misclassification. Hence, the proposed framework provides a valuable analytical tool for hardware-aware robustness evaluation and for developing design automation flows that integrate adversarial stress testing during neuromorphic chip validation.

\section{Conclusion and Future Work}


In this paper, we propose Spike-PTSD, a bio-plausible AEA inspired by PTSD-affected brain dynamics. By simulating abnormal neuronal activations in response to post-traumatic stress, Spike-PTSD controls corresponding spike scaling in victim SNNs to refine the traditional adversarial objective, making it more consistent with SNNs. Extensive experiments demonstrate that Spike-PTSD achieves state-of-the-art performance against rate-coding SNNs, even under targeted settings and temporally encoded data inputs. Moreover, Spike-PTSD can serve as a trigger generation mechanism for constructing stealthy Trojan attacks on SNNs, which will be further explored in our future work. We also plan to integrate Spike-PTSD into neuromorphic design automation frameworks for security validation, enabling automated detection of hardware-level vulnerabilities and adversarial sensitivity in neuromorphic chips.

\begin{acks}
This work was supported in part by the National Natural Science Foundation of China under Grant 62372087 and Grant 62572103, in part by the Research Fund of State Key Laboratory of Processors under Grant CLQ202310, and in part by the China Scholarship Council under Grant 202406070152.
\end{acks}

\bibliographystyle{ACM-Reference-Format}
\bibliography{dac_ref}

\end{document}